\documentclass[letter]{aa} 

\usepackage{graphicx}
\usepackage{txfonts}
\begin{document}

   \title{Imaging the dust sublimation front of a circumbinary disk
   \thanks{Based on observations made with ESO Telescopes at the La Silla Paranal Observatory under program ID 094.D-0865.}}
   \titlerunning{PIONIER image of IRAS08544-4431}
   \author{M. Hillen \inst{1}
          \and
          J. Kluska \inst{2} 
          \and
          J.-B. Le Bouquin \inst{3}
          \and
          H. Van Winckel \inst{1}
          \and
          J.-P. Berger \inst{4}
          \and
          D. Kamath \inst{1}
          \and
          V. Bujarrabal \inst{5}
          }
   \institute{Instituut voor Sterrenkunde (IvS), KU Leuven,
              Celestijnenlaan 200D, B-3001 Leuven, Belgium\\
              \email{Michel.Hillen@ster.kuleuven.be}
         \and
             School of Physics, University of Exeter, Stocker Road, Exeter EX4 4QL, UK
         \and
             UJF-Grenoble 1/CNRS-INSU, Institut de Plan\'etologie et d'Astrophysique de Grenoble (IPAG) UMR 5274, 38041 Grenoble, France
         \and
             ESO, Karl-Schwarzschild-Strasse 2, 85748 Garching bei Munchen, Germany
         \and
             Observatorio Astron\'omico Nacional (OAN-IGN), Apartado 112, 28803 Alcal\'a de Henares, Spain
             }

   \date{Received January 14, 2016; accepted February 16, 2016}
   \authorrunning{Hillen et al.}
   
 
  \abstract
   {}
   {We present the first near-IR milli-arcsecond-scale image of a post-AGB binary that is surrounded by hot circumbinary dust.}
   {A very rich interferometric data set in six spectral channels was acquired of IRAS08544-4431 
   with the new RAPID camera on the PIONIER beam combiner at the Very Large Telescope Interferometer (VLTI). 
   A broadband image in the \textit{H} band was reconstructed by combining the data of all spectral channels using the SPARCO method.
   }
   {We spatially separate all the building blocks of the IRAS08544-4431 system in our milliarcsecond-resolution image. 
   Our dissection reveals a dust sublimation front that is strikingly similar to that expected in early-stage protoplanetary disks, 
   as well as an unexpected flux signal of $\sim$4\% from the secondary star. The energy output from this companion 
   indicates the presence of a compact circum-companion accretion disk, which is likely the origin of the fast outflow detected in H$\alpha$.
   }   
   {Our image provides the most detailed view into the heart of a dusty circumstellar disk to date.
   Our results demonstrate that binary evolution processes and circumstellar disk evolution can be studied in detail in space and over time.   
   }

   \maketitle
%

\section{Introduction}

Binary interactions play a fundamental role in many poorly understood stellar phenomena.
One peculiar class of objects concerns the post-Asymptotic Giant
Branch (post-AGB) stars in SB1 binary systems, which have hot as well as cold circumstellar dust and 
gas \citep{2003ARAAVanWinckel}. 
The presence of a near-IR excess in the spectral energy distribution (SED) 
of a post-AGB star correlates well with the central star 
being part of a $\sim$1-2~au-wide binary system \citep[e.g.][]{2009AAVanWinckel}.
Such evolved binaries are common in the Galaxy \citep{2006AAdeRuyter} and recent studies 
show that about 30\% of all optically bright post-AGB stars have this 
typical SED \citep[][]{2015MNRASKamath}.
The companions are not detected and are assumed to be unevolved, and of low luminosity compared 
to the post-AGB star.

The specific SED of these objects indicates the presence of a stable, circumbinary dust 
reservoir starting at the dust sublimation radius. The single-dish CO line survey of \citet{2013AABujarrabalB} also 
shows their gas structures to be in Keplerian rotation, based on the narrow 
emission profiles of the CO rotationally excited lines. The orbiting gas
in the outer disk has also been spatially resolved in two objects \citep{2015AABujarrabal,2013AABujarrabalC}.

The gas and dust rich disk is passively heated, and therefore vertically puffed-up, by the energy it intercepts from the 
luminous but low-mass post-AGB star. The mid-IR dust emission features and the mm slopes in the SEDs reveal a high degree 
of grain processing and growth \citep[up to mm sizes,][]{2015AAHillen,2011AAGielen}.
To probe the physical and chemical characteristics of this circumstellar material, which is at astronomical 
unit scales in these distant ($\sim$kpc) objects, observations are required with the angular
resolution of a long-baseline interferometer. Several objects have been resolved in the near- or 
mid-IR in this way, confirming their disk nature \citep[][]{2015AAHillen,2007AADeroo}.
In contrast to the protoplanetary disks around young stars \citep{2011AABenisty}, no post-AGB inner disk rim 
has yet been imaged.
The target of this study, IRAS08544-4431, was resolved with a limited baseline 
coverage \citep{2007AADeroo} on the Very Large Telescope Interferometer (VLTI). Here we present 
the first near-IR milliarcsecond-scale image that fully dissects the inner object into its constituent components.


\section{Image reconstruction strategy} \label{section:strategy}
Interferometric image reconstruction is an ill-posed inverse problem aimed at determining image pixel values. 
It is typically solved with a Bayesian approach: a global cost function
$\zeta = \zeta_{data} + \mu \zeta_{rgl}$ 
is minimized, with $\zeta_{data}$ the likelihood term (the $\chi^2$), $\zeta_{rgl}$ a regularization 
term, and $\mu$ the regularization weight. The regularization helps 
to fill the gaps in the UV-coverage by interpolating the Fourier plane in a specific way, 
and is crucial when converging to the most likely estimate of the true source brightness distribution.

Here we produce images with a chromatic reconstruction algorithm, named SPARCO \citep{2014AAKluska}. This method decomposes 
the source brightness distribution in two: a fraction of the flux is included in the form of a 
parametric model, while a model-independent reconstruction is done of the remaining flux (i.e., \emph{the environment}). The total 
complex visibility is then a linear combination of the visibility of each component, weighted by the flux ratio that bears the chromatic 
information:
\begin{equation} \label{eq:decomposition}
 V_{\mathrm{tot},\lambda}(u,v) = \frac{\left[\displaystyle{\sum_i f_{i} \Lambda_{i,\lambda} V_{i,\lambda}(u,v)}\right] + f_{\mathrm{env}} \Lambda_{\mathrm{env},\lambda} V_{\mathrm{env},\lambda}(u,v)}{\left[\displaystyle{\sum_i f_{i} \Lambda_{i,\lambda}}\right] + f_{\mathrm{env}} \Lambda_{\mathrm{env},\lambda}}
\end{equation}
The chromaticity is parameterized by the coefficients $f$ and $\Lambda$ 
($\sum_i f_{i} + f_{env} = 1$, $\forall i : \, \Lambda_{i,1.65} = 1$ and $\Lambda_{env,1.65} = 1$). The chromatic and model parameters can be fixed from prior
information, or be fitted with the image of the environment.

SPARCO uses existing (monochromatic) reconstruction algorithms, in this case MiRA \citep{2008SPIEThiebaut}. 
Our image contains 512x512 pixels, with a pixel size of 0.15 mas/pixel. We apply the quadratic smoothness regularization \citep{2011AARenard}. 
The regularization weight $\mu$ is re-determined for each reconstruction with the L-curve method \citep{2011AARenard}, taking a grid 
of 100 values between 10$^5$ and 10$^{14}$. The uncertainties on the input data are translated into a 
measure of the significance of each pixel value with a bootstrap method \citep{boot}. 
We generate 500 new data sets, of equal size to the original, 
by randomly picking squared-visibility and closure-phase measurements. An image is reconstructed for each virtual data set. 
From this image cube we compute the average and standard deviation image and significance contours at 5, 3, and 1$\sigma$. 

In this paper, we first reconstruct an image with the parametric component equal to a single star, the primary 
(Sect.~\ref{section:imagereconstruction}). Then, we 
fit a model to quantify the detected structures (Sect.~\ref{section:modeling}), and we feed some of the resulting information 
into another reconstruction in which two stellar components are subtracted.

\section{Observations} \label{section:observations}

Interferometric data were obtained with the PIONIER instrument \citep[][]{2011AALeBouquin}, 
combining the 1.8m Auxiliary Telescopes of the VLTI. 
PIONIER was upgraded with a revolutionary fast and low-noise infrared camera \citep[RAPID,][]{2014SPIEGuieu}. 
Table~\ref{tab:obs} summarizes the log of the observations. Unfortunately, one VLTI delay line suffered from technical 
problems during our last run. To calibrate the fringe visibilities and closure phases, we interleaved science observations 
with those of KIII reference stars that were found with 
the \texttt{SearchCal}\footnote{http://www.jmmc.fr/searchcal\_page} software. 
Data were reduced and calibrated with the \texttt{pndrs} package \citep[][]{2011AALeBouquin}. 
The five consecutive measurements in each observing block (OB) were averaged together into a single observation, each 
containing six visibilities and four closure phases dispersed over six spectral channels across the H-band. The resulting 
calibrated observations and uv-coverage are shown in Fig.~\ref{figure:V2data}, and in Figs.~\ref{figure:CPdata} 
and~\ref{figure:uvcoverage} in the appendices.

\begin{table}[h]
\caption{Log of the PIONIER observations.}
\centering
\begin{tabular*}{0.95\columnwidth}{llrrr}
\hline\hline\noalign{\smallskip}
Night & Conf. used & \# OBs & \# V$^2$ & \# CP \\
\noalign{\smallskip}\hline\noalign{\smallskip}
2015-01-21\tablefootmark{a,b,c,d} & D0-G1-H0-I1 & 9 & 324 & 162 \\
2015-01-24\tablefootmark{a,b,c,d,e} & A1-G1-K0-I1 & 10 & 360 & 180 \\
2015-02-23 \tablefootmark{a,b,d,e}& B2-C1-D0 & 8 & 144 & 48 \\
\noalign{\smallskip}\hline
\end{tabular*}
\tablefoot{Calibrators:
\tablefoottext{a}{HD76160 (KIII, $0.47\pm0.1$mas)},
\tablefoottext{b}{HD73075 (KIII, $0.47\pm0.1$mas)},
\tablefoottext{c}{HD77140 (Am, $0.46\pm0.1$mas)}, 
\tablefoottext{d}{HD76111 (KIII, $0.41\pm0.08$mas)}, 
\tablefoottext{e}{HD75104 (KIII, $0.57\pm0.12$mas)}
}
\label{tab:obs}
\end{table}

\begin{figure}
   \centering
   \includegraphics[width=8cm]{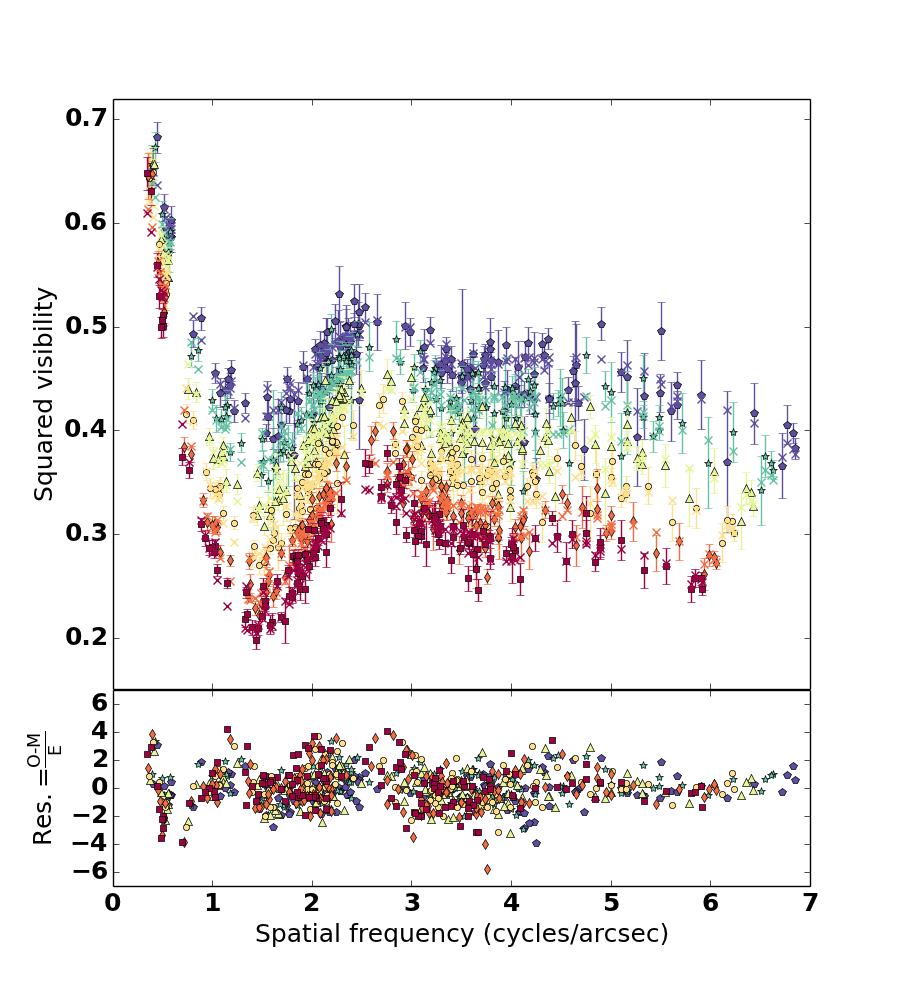}
   \caption{The PIONIER squared visibility data with the best-fit parametric model (upper panel), 
   and normalized residuals (lower panel), as a function of radial spatial frequency. The color and symbol indicate 
   the wavelength in $\mu$m (blue pentagon: 1.53, cyan star: 1.58, green triangle: 1.63, yellow circle: 1.68, 
   orange diamond: 1.73, red square: 1.77). The cross symbols in the upper panel represent model values.}
   \label{figure:V2data}
\end{figure} 

\onlfig{
\begin{figure}
   \centering
   \includegraphics[width=8cm]{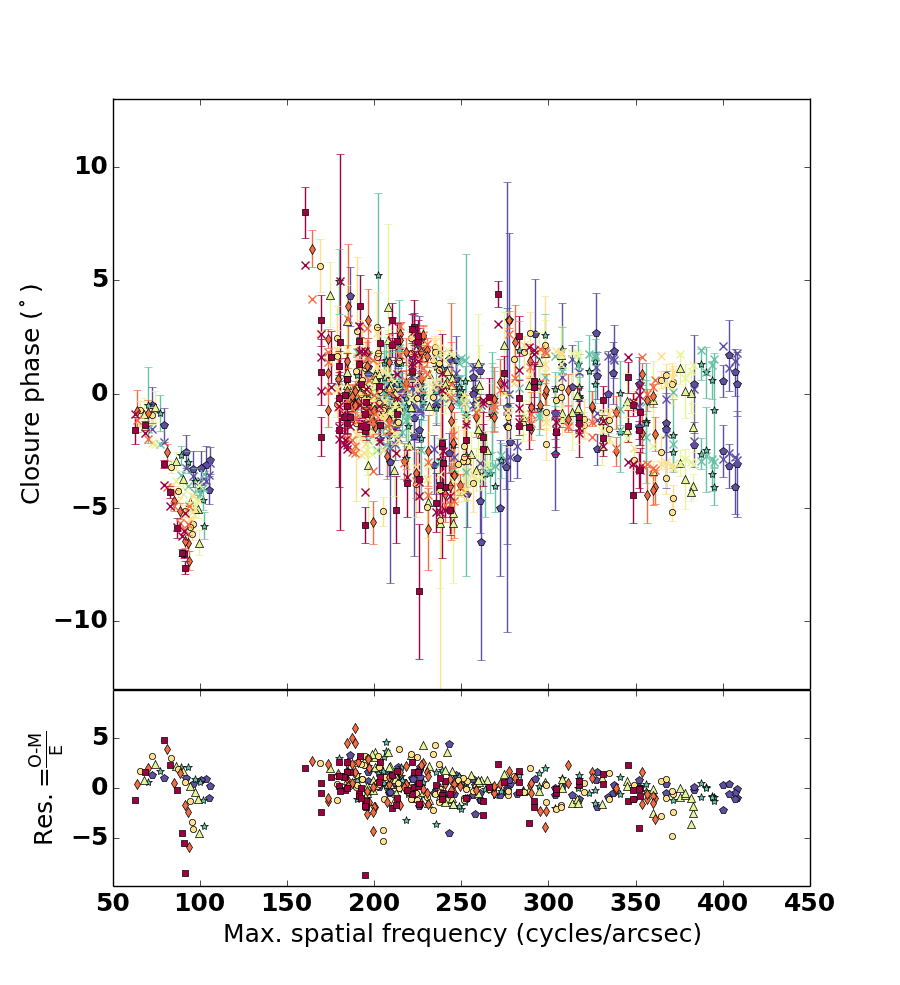}
   \caption{The PIONIER closure phase data as a function of the maximum radial spatial frequency in the given baseline triangle, with the 
   best-fit parametric model (upper panel) and residuals (lower panel). The color and symbol indicate 
   the wavelength in $\mu$m (blue pentagon: 1.53, cyan star: 1.58, green triangle: 1.63, yellow circle: 1.68, 
   orange diamond: 1.73, red square: 1.77).}
   \label{figure:CPdata}
\end{figure} 
}
\onlfig{
\begin{figure}
   \centering
   \includegraphics[width=7cm]{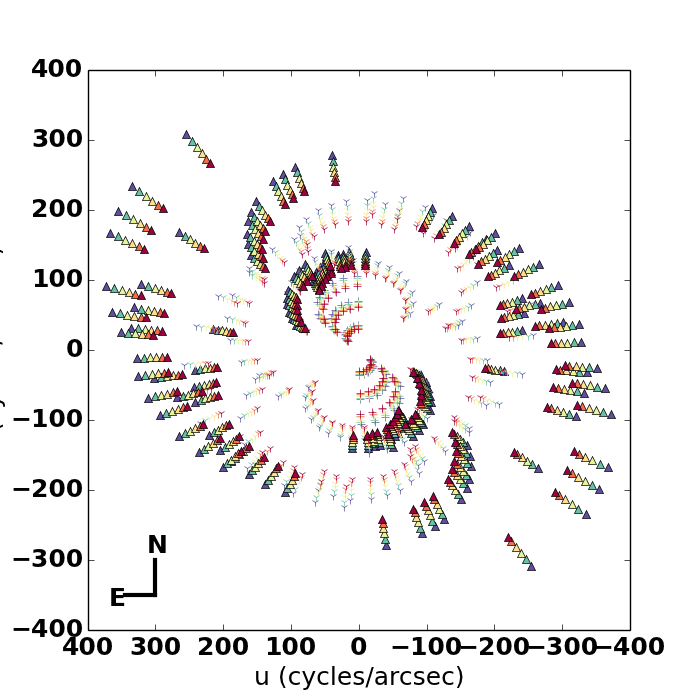}
   \caption{A graphical representation of the uv-coverage. Data taken during the first, second, and third run are indicated with
   tripod, triangle, and plus symbols, respectively. The wavelength of an observation is indicated by the
   color of the symbol, going from 1.53 $\mu$m (blue) to 1.77 $\mu$m (red).}
   \label{figure:uvcoverage}
\end{figure} 
}


\section{Single-star-subtracted reconstruction} \label{section:imagereconstruction}
The visibility of the primary is represented with a uniform disk model (angular diameter $\theta_\star=0.5$~mas) that is fixed to the origin of the 
coordinate system. The chromaticity of the stellar and environment fluxes is well represented by power laws:
$\Lambda_{\mathrm{pri},\lambda} = \left(\frac{\lambda}{1.65 \mu \mathrm{m}}\right)^{-4}$ and
 $\Lambda_{\mathrm{env},\lambda} = \left(\frac{\lambda}{1.65 \mu \mathrm{m}}\right)^{d_{\mathrm{env}}} , \quad f_{\mathrm{env}} = (1-f_{\mathrm{pri}})$
with $d_{env}$ the spectral index of the environment.

We make a 20x20 grid on the chromatic parameters in which 
$f_{\mathrm{pri}}$ ranges from 0.5 to 0.65 and $d_{\mathrm{env}}$ from -3 to 2. 
The posterior probability distribution strongly peaks at $f_{\mathrm{pri}}=0.61$ and $d_{\mathrm{env}}=0.42$. 
For the regularization weight, we use $\mu=10^{12}$. 

\begin{figure}
   \centering
   \includegraphics[width=9cm]{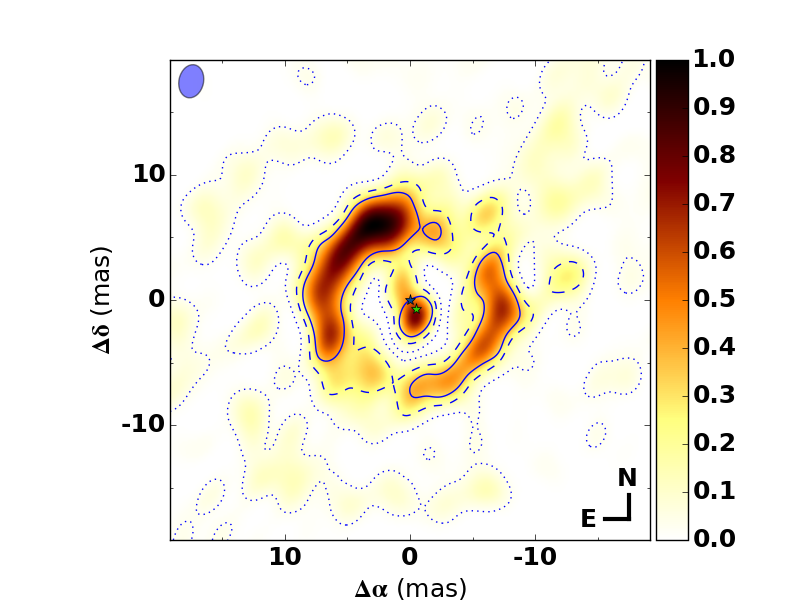}
   \includegraphics[width=9cm]{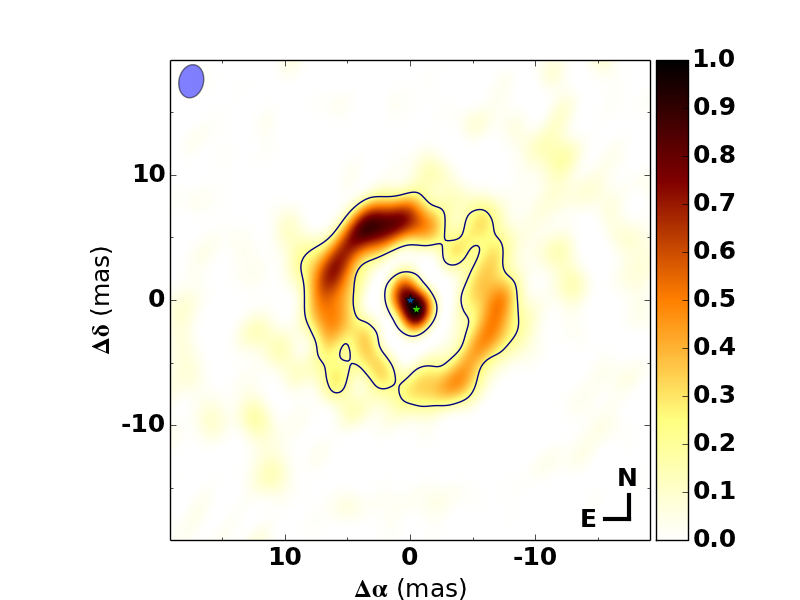} 
   \includegraphics[width=9cm]{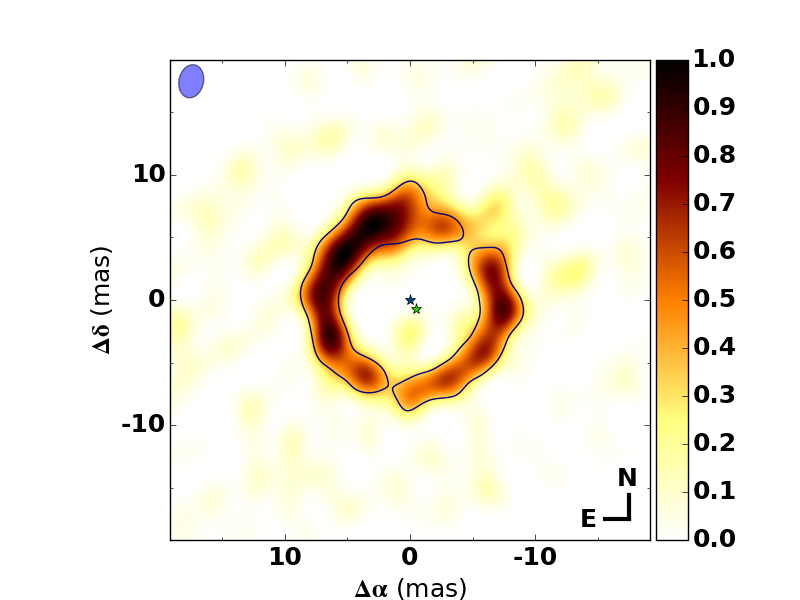}
   \caption{Reconstructed images. Upper panel: single-star-subtracted reconstruction of the original data. Middle panel: single-star-subtracted 
   construction of the synthetic data set made with the best-fit parametric model. Lower panel: binary-subtracted reconstruction of the original data.
   The horizontal and vertical axes contain right ascension and declination coordinates in units of milli-arcseconds, 
   respectively (the north and east directions are indicated in the lower right corner). The blue 
   ellipse shows the beamsize. The full, dashed and dotted lines delineate pixels that have fluxes above 
   5, 3 and 1$\sigma$, respectively. For clarity only the 5$\sigma$ contours are shown in the lower panels. 
   The blue and green star symbols indicate the positions of the two stars 
   as determined from the best-fit parametric model (the actual flux of the primary is subtracted from the image). 
   The color indicates normalized flux, scaled to the highest pixel value in the image.}
   \label{figure:image}
\end{figure}

The upper panel in Fig.~\ref{figure:image} shows the inner 256x256 pixels of the reconstructed image, 
which has a beam size of $\sim$1.3~mas (defined as twice the Gaussian FWHM fitted to the interferometric point spread function). 
Various features can be identified: An inclined but almost circular ring, centered around the primary, is well resolved 
with a detection threshold better than 5$\sigma$. The ring is not uniform in intensity, is $\sim$15~mas in diameter, and appears to be clumpy.
The ring contains two opposite intensity maxima, of which the brightest one is in the northeast direction. Two 
flux minima (close to flux nulls) appear in the northwest and southeast parts of the ring. Additionally, 
there is an unresolved (point-like) emission component at 5$\sigma$ that is offset by $\sim$1~mas from the center in the southwest direction, 
and which extends to the northeast in a faint emission stream (at ~3$\sigma$). Finally, the faint emission on scales larger 
than the ring is well detected when spatially integrated, but its morphology is unconstrained (at $\sim$1$\sigma$).

\section{Parametric modeling} \label{section:modeling}
In this section, we quantify the morphologies and wavelength-dependent flux contributions of the different components, by applying
increasingly complex parametric models. 

We perform the model-fitting in multiple steps, subsequently adding components to the system. 
The primary star is represented in the same way as in the first image reconstruction. 
We add: 1) an inclined ring with a 
radial profile that is described by a Gaussian function and which can vary in brightness as a function 
of azimuthal angle with an m=1 and m=2 modulation (Model 1 in Table~\ref{table:onlineparametric} in the appendix), 2) a uniform background flux that fills 
the whole field-of-view (Model 2 in Table~\ref{table:onlineparametric}), and 3) an unresolved point source (interpreted as the companion star), 
slightly offset from the primary (\emph{Best model} in Table~\ref{table:parametric}).
The ring is mathematically described in \citet{2012SPIEKluska}. We attach 
the origin of the coordinate system to the center of the ring. 
The background flux is completely resolved, irrespective of the baseline. 
The positions of the two stars are defined by the projected separation vector's length 
and position angle, because we anchor the center of mass of the binary system to the center of the ring. 
For this we assume co-planarity between the orbital plane and the disk midplane. 
Then we use the binary mass ratio -- from the spectroscopic mass function \citep{2003AAMaas} and a typical post-AGB mass of 
$\sim$0.6~M$_\odot$ for the primary -- to compute the position of the center of mass along the projected separation vector.

For the ring, background, and secondary star components, we define functions $\Lambda_{i,\lambda}$ as normalized black bodies 
with temperatures T$_\mathrm{r}$, T$_{\mathrm{back}}$ and T$_{\mathrm{sec}}$, respectively. The $\Lambda_{\mathrm{pri},\lambda}$ function 
is represented with a Kurucz atmosphere model \citep[$T_{\mathrm{eff}}=7250$~K, $\log g = 1.0$, \lbrack Fe/H\rbrack=-0.5,][]{2003AAMaas} that
is convolved to the spectral resolution of PIONIER.

We use a Markov Chain Monte Carlo (MCMC) method in a Bayesian statistics framework to estimate the posterior probabilities 
of the 16 parameters in our model. We define 
uniform priors, the ranges of which are determined on the basis of the reconstructed image 
(e.g., the ring diameter and inclination), physically acceptable ranges (e.g., the binary separation), 
and previous knowledge of the system (e.g., the primary and secondary's fluxes). The likelihood is implemented as 
$\exp(-\chi^2/2)$, with $\chi^2$ the standard goodness-of-fit parameter. We apply the ensemble sampler 
with affine-invariance in the emcee python package \citep{2013PASPForeman}. We 
experiment with the number of chains and the number of steps per chain, and find 
reproducible results with 400 chains and 1000 steps. The best-fit parameter values and their uncertainties 
are computed as the 16th, 50th, and 84th percentiles of the samples in the marginalized distributions, but we
quote the average of the upper and lower bounds as a single uncertainty. The assumption of 
Gaussian uncorrelated noise is probably not entirely justified. 
Thanks to our dedicated observing strategy, the dominant source of correlated noise is likely to stem from the 
simultaneity of the measurements in the six spectral channels. We take this into account 
by multiplying the data uncertainties by a factor $\sqrt{6}$ in the MCMC. 

\begin{table}
 \caption{Summary of the parametric modeling. Only the best-fit model is included. The results of intermediate modeling steps 
 are shown in Table~\ref{table:onlineparametric} in the appendix.}             
 \label{table:parametric}      
 \centering                          
 \begin{tabular}{l l r @{ -- } l r @{ $\pm$ } l}        
 \hline\hline\noalign{\smallskip}
        & Parameter & \multicolumn{2}{c}{Prior} & \multicolumn{2}{c}{Best model} \\
\noalign{\smallskip}\hline\noalign{\smallskip}
Primary & $\theta_{\mathrm{pri}}$ (mas) & \multicolumn{2}{c}{0.5} & \multicolumn{2}{c}{0.5} \\
        & $f_{\mathrm{pri}}$ & 0.3 & 0.8 & 0.597 & 0.006 \\
Stellar positions & $\rho_{\mathrm{bin}}$ (mas) & 0.0 & 5.0 & 0.81 & 0.05 \\
                  & PA$_{\mathrm{bin}}$ ($^\circ$) & 0 & 360 & 56 & 3 \\
Ring & $\Theta$ (mas) & 2 & 20 & 14.15 & 0.10 \\
     & $\delta \Theta / \Theta$ & 0.0 & 1.0 & 0.457 & 0.015  \\
     & c$_1$ & -1.0 & 1.0 & 0.35 & 0.04 \\
     & s$_1$ & -1.0 & 1.0 & -0.25 & 0.07 \\
     & c$_2$ & -1.0 & 1.0 & -0.05 & 0.06 \\
     & s$_2$ & -1.0 & 1.0 & -0.27 & 0.03 \\
     & PA ($^\circ$) & 0 & 360 & 6 & 6 \\
     & i ($^\circ$) & 5 & 50 & 19 & 2 \\
     & $f_{\mathrm{r}}$ & 0.2 & 0.7 & 0.209 & 0.009 \\
     & T$_\mathrm{r}$ (K) & 500 & 3000 & 1120 & 50 \\
Background  & $f_{\mathrm{back}}$ & 0.0 & 0.5 & 0.155 & 0.005 \\
            & T$_{\mathrm{back}}$ (K) & 500 & 8000 & 2400 & 300 \\
Secondary   & $\theta_{\mathrm{sec}}$ (mas) & \multicolumn{2}{c}{0.0} & \multicolumn{2}{c}{0.0} \\
            & $f_{\mathrm{sec}}$ & 0.0 & 0.2 & 0.039 & 0.007 \\
            & T$_{\mathrm{sec}}$ (K) & 1500 & 8000 & 4000 & 2000 \\
\noalign{\smallskip}\hline\noalign{\smallskip}            
Chi-square  & $\chi^2_r[\mathrm{V}^2]$ & & & \multicolumn{2}{c}{1.8} \\
            & $\chi^2_r[\mathrm{CP}]$ & & & \multicolumn{2}{c}{3.3} \\
            & $\chi^2_r[\mathrm{all}]$ & & & \multicolumn{2}{c}{2.3} \\
\noalign{\smallskip}\hline
\end{tabular}
\end{table}

\onltab{
\begin{table*}
 \caption{Summary of the intermediate steps in the parametric modeling.}             
 \label{table:onlineparametric}      
 \centering                          
 \begin{tabular}{l l r @{ -- } l r @{ $\pm$ } l r @{ $\pm$ } l}        
 \hline\hline\noalign{\smallskip}
        & Parameter & \multicolumn{2}{c}{Prior} & \multicolumn{2}{c}{Model1} & \multicolumn{2}{c}{Model2} \\
\noalign{\smallskip}\hline\noalign{\smallskip}
Primary & $\theta_{\mathrm{pri}}$ (mas) & \multicolumn{2}{c}{0.5} & \multicolumn{2}{c}{0.5} & \multicolumn{2}{c}{0.5} \\
        & $f_{\mathrm{pri}}$ & 0.3 & 0.8 & 0.625 & 0.001 & 0.628 & 0.001 \\
Stellar positions & $\rho_{\mathrm{bin}}$ (mas) & 0.0 & 5.0 & \multicolumn{2}{c}{0.0} & \multicolumn{2}{c}{0.0} \\
                  & PA$_{\mathrm{bin}}$ ($^\circ$) & 0 & 360 & \multicolumn{2}{c}{ -- } & \multicolumn{2}{c}{ -- } \\
Ring & $\Theta$ (mas) & 2 & 20 & 14.13 & 0.15 & 14.25 & 0.16 \\
     & $\delta \Theta / \Theta$ & 0.0 & 1.0 & 0.99 & 0.01 & 0.48 & 0.02  \\
     & c$_1$ & -1.0 & 1.0 & 0.25 & 0.02 & 0.18 & 0.02 \\
     & s$_1$ & -1.0 & 1.0 & 0.18 & 0.04 & -0.20 & 0.04 \\
     & c$_2$ & -1.0 & 1.0 & 0.20 & 0.08 & -0.20 & 0.04 \\
     & s$_2$ & -1.0 & 1.0 & -0.25 & 0.07 & -0.09 & 0.05 \\
     & PA ($^\circ$) & 0 & 360 & 13 & 7 & 345 & 5 \\
     & i ($^\circ$) & 5 & 50 & 27 & 3 & 25 & 2 \\
     & $f_{\mathrm{r}}$ & 0.2 & 0.7 & 0.375 & 0.001 & 0.208 & 0.005 \\
     & T$_\mathrm{r}$ (K) & 500 & 3000 & 1450 & 20 & 1230 & 60 \\
Background  & $f_{\mathrm{back}}$ & 0.0 & 0.5 & \multicolumn{2}{c}{ -- } & 0.164 & 0.005 \\
            & T$_{\mathrm{back}}$ (K) & 500 & 8000 & \multicolumn{2}{c}{ -- } & 2000 & 200 \\
Secondary   & $\theta_{\mathrm{sec}}$ (mas) & \multicolumn{2}{c}{0.0} & \multicolumn{2}{c}{ -- } & \multicolumn{2}{c}{ -- } \\
            & $f_{\mathrm{sec}}$ & 0.0 & 0.2 & \multicolumn{2}{c}{ -- } & \multicolumn{2}{c}{ -- } \\
            & T$_{\mathrm{sec}}$ (K) & 1500 & 8000 & \multicolumn{2}{c}{ -- } & \multicolumn{2}{c}{ -- } \\
\noalign{\smallskip}\hline\noalign{\smallskip}            
Chi-square  & $\chi^2_r[V^2]$ & & & \multicolumn{2}{c}{9.0} & \multicolumn{2}{c}{3.0} \\
            & $\chi^2_r[CP]$ & & & \multicolumn{2}{c}{6.9} & \multicolumn{2}{c}{6.7} \\
            & $\chi^2_r[\text{all}]$ & & & \multicolumn{2}{c}{8.2} & \multicolumn{2}{c}{4.4} \\
\noalign{\smallskip}\hline
\end{tabular}
\end{table*}
}

The squared visibilities and closure phases of the best-fit model are included in Fig.~\ref{figure:V2data} and online Fig.~\ref{figure:CPdata}, 
along with the original data and the residuals. Most model parameters are well constrained 
(Table~\ref{table:parametric}, and Table~\ref{table:onlineparametric} and Fig.~\ref{figure:MCMCcornerplot} in the appendix). 
With each geometric component that is added, 
the total reduced chi-square decreases by a factor of two, with the final model having $\chi^2_r \sim 2.3$. 
The residuals are dominated by certain closure phase measurements at low and intermediate spatial 
frequencies, which have very small uncertainties. Our fairly simple parameterization of the ring and background flux 
may not fully capture the asymmetry probed by these precise data. 

\onlfig{
\begin{figure*}
   \centering
   \includegraphics[width=18cm]{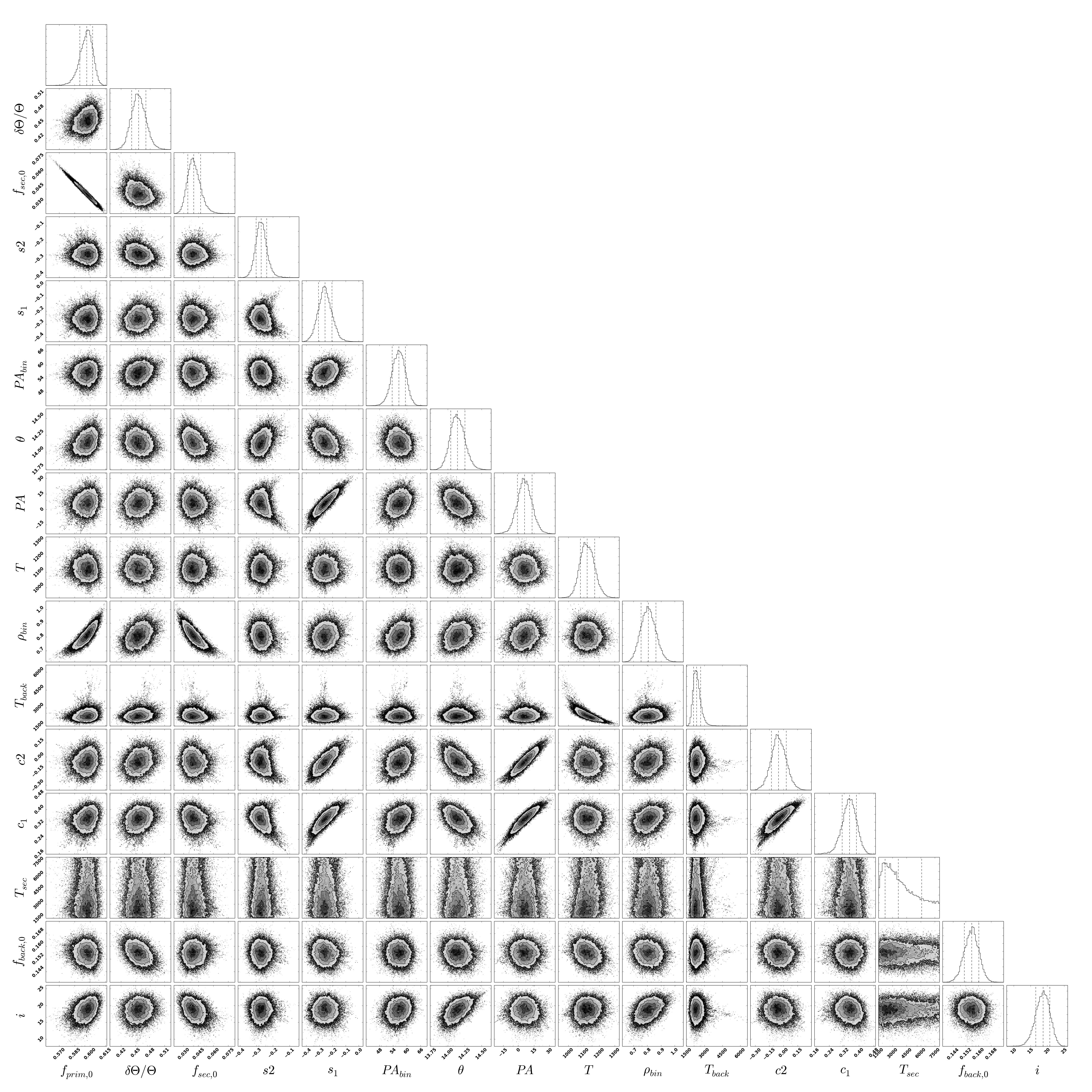}
   \caption{The MCMC ``corner plot'' of our best-fit parametric model (Model3). It shows 
   the one- and two-dimensional projections of the posterior probability distributions of all 
   the parameters. The histograms on the diagonal also include the positions of the 16th, 50th, and 84th percentiles.}
   \label{figure:MCMCcornerplot}
\end{figure*} 
}

Most of the flux at 1.65~$\mu$m is emitted by the primary star (59.7$\pm$0.6\%), followed by the ring (20.9$\pm$0.5\%), the 
over-resolved background (15.5$\pm$0.5\%), and the companion star (3.9$\pm$0.7\%).  Their best-fit temperatures are 7250~K (fixed), 
1120$\pm$50~K, 2400$\pm$300~K, and 4000$\pm$2000~K, respectively.  
The angular separation between the two stellar components is $\rho=0.81\pm0.05$~mas. The diameter of the ring is $\Theta=14.15\pm0.10$~mas, 
about 18 times the binary separation, and has a Gaussian width of $\mathrm{FWHM}=3.2\pm0.1$~mas. The ring is inclined 
with respect to the plane of the sky by $i=19\pm2^\circ$. A 
model image of the ring is included in Fig.~\ref{figure:modelimage} in the appendix.

\onlfig{
\begin{figure}
   \centering
    \includegraphics[width=9cm]{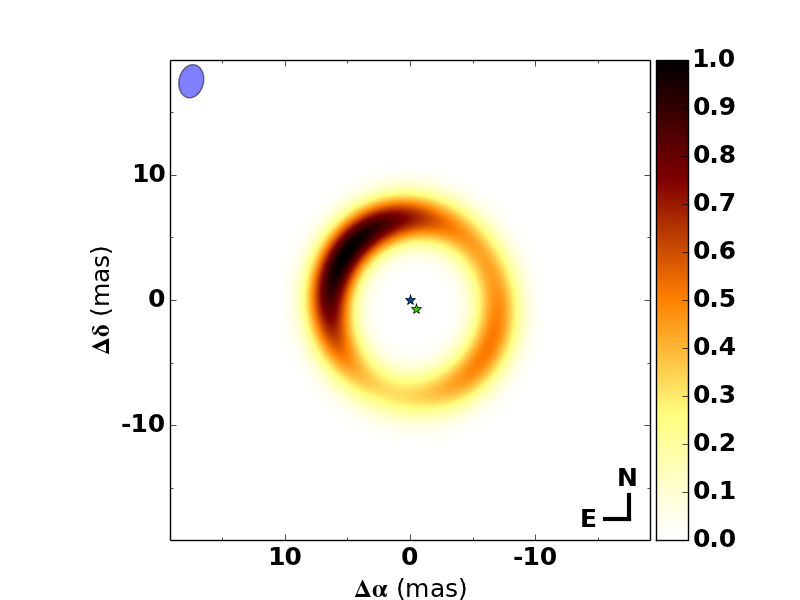}
   \caption{The theoretical brightness distribution of our best-fit parametric model. The color bar shows normalized flux, 
   scaled to the highest pixel value in the image. The position of the primary and secondary star are 
   indicated with a blue and green star symbol, respectively.}
   \label{figure:modelimage}
\end{figure} 
}

As a consistency check, we reconstruct an image from a synthetic data set of the best-fit model, 
using the same image parameters (number of pixels, pixel sizes) and regularization (both type and weight) as before. 
The reconstruction of the model bears strong resemblance to the reconstruction of the real data (see Fig.~\ref{figure:image}). 
The similarity in the background flux distribution shows its shaping is a direct consequence of the UV-coverage. 
The secondary is present as well, but there is residual flux at the location of the primary. 
As this residual flux should not be present, we do another reconstruction of the real data in which the binary system is subtracted
instead of only the primary. The sum within square brackets in Eq.~\ref{eq:decomposition} is 
replaced by the best-fit binary parameters listed in Table~\ref{table:parametric}. The result is shown 
in the lower panel of Fig.~\ref{figure:image}. 
The lack of residual emission in the center of this image validates our parametric model results and shows that there are 
artifacts that are due to the reconstruction process in the single-star-subtracted images (e.g., the northeast emission stream inside the ring).
The disk also emerges more prominently as a ring with reduced intensity asymmetries.


\section{Discussion}
Our results provide the first direct view into the central region of an evolved binary surrounded by a circumbinary disk.

The ring can be readily associated with the inner dust sublimation rim of the circumbinary disk, which is 
resolved at an unprecedented relative scale (>5 physical resolution elements). Its temperature of 
$\sim$1150~K is consistent with the typical sublimation temperature of silicate grains \citep{2009AAKama}. 

IRAS08544-4431 is the first post-AGB binary system in 
which direct emission from the secondary is detected, and even spatially separated 
from the primary. There are two hypotheses to explain the high companion flux at 1.65 $\mu$m: thermal emission from the 
surface of a 1.5-2.0~M$_\odot$ red giant or emission from a compact accretion disk around a 1.5-2.0~M$_\odot$ main-sequence star. 
We consider the second case more likely because observations of similar post-AGB systems
indicate that circumcompanion accretion disks may be common \citep{2015MNRASGorlova,2012AAGorlova}. The main 
evidence comes from the detection (in H$\alpha$ and for more inclined systems) of fast outflows that originate from the companion. 
The H$\alpha$ line of IRAS08544-4431 has 
a P Cygni-like profile \citep{2003AAMaas}, which is consistent with this interpretation, given that we find the system 
to be viewed close to face-on.



Our observations demonstrate that the inner dynamics in these evolved systems can now be spatially resolved in real-time.
By doing so over time, a new route is open to study and constrain the complex physical processes that 
govern disk evolution and dispersal around evolved binaries.

\begin{acknowledgements}
MH and HvW acknowledge support from the Research Council of the KU Leuven under grant number GOA/2013/012. 
JK acknowledges support from a Marie Curie Career Integration Grant (SH-06192, PI: Stefan Kraus) and the French POLCA project (ANR-10-BLAN-0511).
PIONIER is funded by the Universit\'e Joseph Fourier (UJF), the Institut de Plan\'etologie et d'Astrophysique de Grenoble (IPAG), 
the Agence Nationale pour la Recherche (ANR-06-BLAN-0421, ANR-10-BLAN-0505, ANR-10-LABX56, ANR-11-LABX-13), and the 
Institut National des Science de l'Univers (INSU PNP and PNPS). The integrated optics beam combiner is the result of 
a collaboration between IPAG and CEA-LETI based on CNES R\&T funding.  This research has made use 
of the JMMC SearchCal service (available at http://www.jmmc.fr/searchcal), 
co-developped by FIZEAU and LAOG/IPAG.  
\end{acknowledgements}

\bibliographystyle{aa}
\bibliography{ImageIRAS08544.bib}

\begin{thebibliography}{23}
\expandafter\ifx\csname natexlab\endcsname\relax\def\natexlab#1{#1}\fi

\bibitem[{{Benisty} {et~al.}(2011){Benisty}, {Renard}, {Natta}, {Berger},
  {Massi}, {Malbet}, {Garcia}, {Isella}, {M{\'e}rand}, {Monin}, {Testi},
  {Thi{\'e}baut}, {Vannier}, \& {Weigelt}}]{2011AABenisty}
{Benisty}, M., {Renard}, S., {Natta}, A., {et~al.} 2011, \aap, 531, A84

\bibitem[{{Bujarrabal} {et~al.}(2013{\natexlab{a}}){Bujarrabal}, {Alcolea},
  {Van Winckel}, {Santander-Garc{\'{\i}}a}, \&
  {Castro-Carrizo}}]{2013AABujarrabalB}
{Bujarrabal}, V., {Alcolea}, J., {Van Winckel}, H., {Santander-Garc{\'{\i}}a},
  M., \& {Castro-Carrizo}, A. 2013{\natexlab{a}}, \aap, 557, A104

\bibitem[{{Bujarrabal} {et~al.}(2015){Bujarrabal}, {Castro-Carrizo}, {Alcolea},
  \& {Van Winckel}}]{2015AABujarrabal}
{Bujarrabal}, V., {Castro-Carrizo}, A., {Alcolea}, J., \& {Van Winckel}, H.
  2015, \aap, 575, L7

\bibitem[{{Bujarrabal} {et~al.}(2013{\natexlab{b}}){Bujarrabal},
  {Castro-Carrizo}, {Alcolea}, {Van Winckel}, {S{\'a}nchez Contreras},
  {Santander-Garc{\'{\i}}a}, {Neri}, \& {Lucas}}]{2013AABujarrabalC}
{Bujarrabal}, V., {Castro-Carrizo}, A., {Alcolea}, J., {et~al.}
  2013{\natexlab{b}}, \aap, 557, L11

\bibitem[{{de Ruyter} {et~al.}(2006){de Ruyter}, {van Winckel}, {Maas}, {Lloyd
  Evans}, {Waters}, \& {Dejonghe}}]{2006AAdeRuyter}
{de Ruyter}, S., {van Winckel}, H., {Maas}, T., {et~al.} 2006, \aap, 448, 641

\bibitem[{{Deroo} {et~al.}(2007){Deroo}, {Acke}, {Verhoelst}, {Dominik},
  {Tatulli}, \& {van Winckel}}]{2007AADeroo}
{Deroo}, P., {Acke}, B., {Verhoelst}, T., {et~al.} 2007, \aap, 474, L45

\bibitem[{Efron \& Tibshirani(1994)}]{boot}
Efron, B. \& Tibshirani, R.~J. 1994, {An introduction to the bootstrap},
  Chapman \& Hall/CRC Monographs on Statistics \& Applied Probability (Taylor
  \& Francis)

\bibitem[{{Foreman-Mackey} {et~al.}(2013){Foreman-Mackey}, {Hogg}, {Lang}, \&
  {Goodman}}]{2013PASPForeman}
{Foreman-Mackey}, D., {Hogg}, D.~W., {Lang}, D., \& {Goodman}, J. 2013, \pasp,
  125, 306

\bibitem[{{Gielen} {et~al.}(2011){Gielen}, {Bouwman}, {van Winckel}, {Lloyd
  Evans}, {Woods}, {Kemper}, {Marengo}, {Meixner}, {Sloan}, \&
  {Tielens}}]{2011AAGielen}
{Gielen}, C., {Bouwman}, J., {van Winckel}, H., {et~al.} 2011, \aap, 533, A99

\bibitem[{{Gorlova} {et~al.}(2012){Gorlova}, {Van Winckel}, {Gielen}, {Raskin},
  {Prins}, {Pessemier}, {Waelkens}, {Fr{\'e}mat}, {Hensberge}, {Dumortier},
  {Jorissen}, \& {Van Eck}}]{2012AAGorlova}
{Gorlova}, N., {Van Winckel}, H., {Gielen}, C., {et~al.} 2012, \aap, 542, A27

\bibitem[{{Gorlova} {et~al.}(2015){Gorlova}, {Van Winckel}, {Ikonnikova},
  {Burlak}, {Komissarova}, {Jorissen}, {Gielen}, {Debosscher}, \&
  {Degroote}}]{2015MNRASGorlova}
{Gorlova}, N., {Van Winckel}, H., {Ikonnikova}, N.~P., {et~al.} 2015, \mnras,
  451, 2462

\bibitem[{{Guieu} {et~al.}(2014){Guieu}, {Feautrier}, {Zins}, {Le Bouquin},
  {Stadler}, {Kern}, {Rothman}, {Tauvy}, {Coussement}, {de Borniol}, {Gach},
  {Jacquard}, {Moulin}, {Rochat}, {Delboulb}, {Derelle}, {Robert},
  {Vuillermet}, {M{\'e}rand}, \& {Bourget}}]{2014SPIEGuieu}
{Guieu}, S., {Feautrier}, P., {Zins}, G., {et~al.} 2014, in Society of
  Photo-Optical Instrumentation Engineers (SPIE) Conference Series, Vol. 9146,
  Society of Photo-Optical Instrumentation Engineers (SPIE) Conference Series,
  1

\bibitem[{{Hillen} {et~al.}(2015){Hillen}, {de Vries}, {Menu}, {Van Winckel},
  {Min}, \& {Mulders}}]{2015AAHillen}
{Hillen}, M., {de Vries}, B.~L., {Menu}, J., {et~al.} 2015, \aap, 578, A40

\bibitem[{{Kama} {et~al.}(2009){Kama}, {Min}, \& {Dominik}}]{2009AAKama}
{Kama}, M., {Min}, M., \& {Dominik}, C. 2009, \aap, 506, 1199

\bibitem[{{Kamath} {et~al.}(2015){Kamath}, {Wood}, \& {Van
  Winckel}}]{2015MNRASKamath}
{Kamath}, D., {Wood}, P.~R., \& {Van Winckel}, H. 2015, \mnras, 454, 1468

\bibitem[{{Kluska} {et~al.}(2014){Kluska}, {Malbet}, {Berger}, {Baron},
  {Lazareff}, {Le Bouquin}, {Monnier}, {Soulez}, \&
  {Thi{\'e}baut}}]{2014AAKluska}
{Kluska}, J., {Malbet}, F., {Berger}, J.-P., {et~al.} 2014, \aap, 564, A80

\bibitem[{{Kluska} {et~al.}(2012){Kluska}, {Malbet}, {Berger}, {Lazareff}, {Le
  Bouquin}, {Benisty}, {Menard}, {Pinte}, {Millan-Gabet}, \&
  {Traub}}]{2012SPIEKluska}
{Kluska}, J., {Malbet}, F., {Berger}, J.-P., {et~al.} 2012, in Society of
  Photo-Optical Instrumentation Engineers (SPIE) Conference Series, Vol. 8445,
  Society of Photo-Optical Instrumentation Engineers (SPIE) Conference Series,
  84450O

\bibitem[{{Le Bouquin} {et~al.}(2011){Le Bouquin}, {Berger}, {Lazareff},
  {Zins}, {Haguenauer}, {Jocou}, {Kern}, {Millan-Gabet}, {Traub}, {Absil},
  {Augereau}, {Benisty}, {Blind}, {Bonfils}, {Bourget}, {Delboulbe},
  {Feautrier}, {Germain}, {Gitton}, {Gillier}, {Kiekebusch}, {Kluska},
  {Knudstrup}, {Labeye}, {Lizon}, {Monin}, {Magnard}, {Malbet}, {Maurel},
  {M{\'e}nard}, {Micallef}, {Michaud}, {Montagnier}, {Morel}, {Moulin},
  {Perraut}, {Popovic}, {Rabou}, {Rochat}, {Rojas}, {Roussel}, {Roux},
  {Stadler}, {Stefl}, {Tatulli}, \& {Ventura}}]{2011AALeBouquin}
{Le Bouquin}, J.-B., {Berger}, J.-P., {Lazareff}, B., {et~al.} 2011, \aap, 535,
  A67

\bibitem[{{Maas} {et~al.}(2003){Maas}, {Van Winckel}, {Lloyd Evans}, {Nyman},
  {Kilkenny}, {Martinez}, {Marang}, \& {van Wyk}}]{2003AAMaas}
{Maas}, T., {Van Winckel}, H., {Lloyd Evans}, T., {et~al.} 2003, \aap, 405, 271

\bibitem[{{Renard} {et~al.}(2011){Renard}, {Thi{\'e}baut}, \&
  {Malbet}}]{2011AARenard}
{Renard}, S., {Thi{\'e}baut}, E., \& {Malbet}, F. 2011, \aap, 533, A64

\bibitem[{{Thi{\'e}baut}(2008)}]{2008SPIEThiebaut}
{Thi{\'e}baut}, E. 2008, in Society of Photo-Optical Instrumentation Engineers
  (SPIE) Conference Series, Vol. 7013, Society of Photo-Optical Instrumentation
  Engineers (SPIE) Conference Series, 70131I

\bibitem[{{van Winckel}(2003)}]{2003ARAAVanWinckel}
{van Winckel}, H. 2003, \araa, 41, 391

\bibitem[{{van Winckel} {et~al.}(2009){van Winckel}, {Lloyd Evans}, {Briquet},
  {De Cat}, {Degroote}, {De Meester}, {De Ridder}, {Deroo}, {Desmet},
  {Drummond}, {Eyer}, {Groenewegen}, {Kolenberg}, {Kilkenny}, {Ladjal},
  {Lefever}, {Maas}, {Marang}, {Martinez}, {{\O}stensen}, {Raskin}, {Reyniers},
  {Royer}, {Saesen}, {Uytterhoeven}, {Vanautgaerden}, {Vandenbussche}, {van
  Wyk}, {Vu{\v c}kovi{\'c}}, {Waelkens}, \& {Zima}}]{2009AAVanWinckel}
{van Winckel}, H., {Lloyd Evans}, T., {Briquet}, M., {et~al.} 2009, \aap, 505,
  1221

\end{thebibliography}




\end{document}